\documentclass[reprint]{revtex4-1}

\usepackage{amsmath} 
\usepackage{lineno}
\usepackage{mathptmx} 
\usepackage{helvet} 
\usepackage{courier} 
\usepackage{type1cm} 
\usepackage{makeidx} 
\usepackage{graphicx} 
\usepackage{caption}
\usepackage{subcaption}
\usepackage{epstopdf} 

\usepackage[bottom]{footmisc}

\makeindex 

\begin{document}


\title{Propensity and stickiness in the naming game - Tipping fractions of minorities}

\author{Andrew M. Thompson$^{1,2}$}
\author{Boleslaw K. Szymanski$^{1,3,4}$}
\author{Chjan C. Lim$^{1,2}$}

\affiliation{$^1$Network Science and Technology Center, Rensselaer Polytechnic Institute, 110 8th Street, Troy, New York 12180-3590, USA 
\\$^2$Department of Mathematics, Rensselaer Polytechnic Institute, 110 8th Street, Troy, New York 12180-3590, USA
\\$^3$Department of Computer Science, Rensselaer Polytechnic Institute, 100 8th Street, Troy, New York 12180-3590, USA
\\$^4$The Faculty of Computer Science and Management, Wroclaw University of Technology, Wyb. Wyspianskiego 27, 50-370 Wroclaw, Poland}


\begin{abstract}
Agent-based models of the binary naming game are generalized here to represent a family of models parameterized by the introduction of two continuous parameters. These parameters define varying listener-speaker interactions on the individual level with one parameter controlling the speaker and the other controlling the listener of each interaction. The major finding presented here is that the generalized naming game preserves the existence of critical thresholds for the size of committed minorities. Above such threshold, a committed minority causes a fast (in time logarithmic in size of the network) convergence to consensus, even when there are other parameters influencing the system. Below such threshold, reaching consensus requires time exponential in the size of the network. Moreover, the two introduced parameters cause bifurcations in the stabilities of the system's fixed points and may lead to changes in the system's consensus. 

\end{abstract}

\maketitle


\section{Introduction}
Opinion dynamics studies in Network Science \cite{1, 2, 3} have been concerned with the effect of social influence. Prior research has shown that the interpersonal relationships in the network have a dominant effect even in presence of outside pressure \cite{4, 5}; such relations manifest themselves as large-scale changes to the opinions in the network. Various models have been proposed to incorporate this effect such as the Voter Model \cite{6}, Threshold Model \cite{7}, Bass Model \cite{8}, and the naming game \cite{9, 10, 11, 12, 13, 14, 15, 16, 17, 18, 19}. The difference between the naming game and other models is that in the naming game an individual may possess more than one opinion at a time resulting in a mixed opinion state. This paper focuses on the naming game.

The naming game (NG) is an opinion based model in which each node (individual) in the network possesses a list of opinions which comprise its opinion state \cite{9,13,14,15,16, 20}. In the version of NG model considered in this paper each interaction starts with randomly selecting the speaker first, and then randomly choosing the listener among the speaker's neighbors. The speaker then shares a randomly chosen opinion from its opinion state with the listener. Various social rules are imposed to determine the outcome of such sharing. They can be classified into three types "Original", "Listener-only", and "Speaker-only" \cite{19}. In all three cases, the listener adds the shared opinion if it does not have it as a part of its state. If it does, in the "Original" version of NG, both the sender and the listener set their state to the single opinion shared in the interaction. For "Listener-only" ("Speaker-only") only the listener (the speaker) does so. This paper focuses on the "Listener-only" variant of the naming game. Although the naming game allows for many opinions and their subsets to define the states of the node, we restrict our attention to the simplest version of the model in which at most two opinions coexist in the network to which we will refer as binary NG model. For this model we study how modified rules of the outcomes of opinion sharing change the dynamics of the NG outcomes.

Several papers study the dynamics of the naming game in respect to the network topology \cite{13,15, 21, 22, 23}. We have chosen an alternative approach in which the network topology is set to be a complete graph \cite{14, 24, 25}, so the intrinsic properties of the naming game can be analyzed. Two of such properties are the global consensus in the network and the average time to reach it \cite{22}. For the simplified version of the naming game on a network with $N>>1$ nodes, when the threshold fraction of committed agents has not been reached, it has been shown by simulations in \cite{24} and by quasi-stationary approximation in \cite{26} that the time to consensus is $T_c \sim e^N$. 

Committed agents in the naming game are nodes in the network holding one opinion which they never alter. In the presence of committed agents \cite{22, 27}, it is possible to achieve consensus on one opinion quickly when the fraction of committed agents is greater than a certain threshold. On a complete graph, this threshold is $p_c \approx 0.0979$ \cite{26, 28}. The consensus time in such case drops to $T_c \sim \ln N$, as shown by an eikonal approximation in \cite{29}. In this paper, the threshold fraction of committed agents needed for fast consensus will be studied under modified rules on the interactions in the naming game. 

Such modified rules have been studied previously by Baronchelli et al. \cite{18} and Strogatz' group \cite{30}, among others. In the former work, Baronchelli et al. considers the "Original" naming game with a parameter $\beta$ controlling the likelihood of the two interacting agents to update their states to the single opinion shared in their interaction. It was shown that a threshold of $\beta = 1/3$ exists such that consensus does not occur below this threshold. In Marvel et al. \cite{30}, the authors attempt to promote a majority of mixed state holders in the network. In the next sections, the proposed family of naming game models is defined and analyzed in terms of the stability of its equilibria.

\section{Mean Field and Stability Analysis for Modified naming game}

In this paper, we consider the binary "Listener-only" naming game in which two unique opinions, $A$ and $B$, are spreading in the network. The evolution of the number of $A$ and $B$ opinion holders can be cast as a random walk in 2D space. 
Instead of working at the micro-state level, we consider the macrostate of the system. Each node in the network holds either one of the {\it unique opinion}, or their conjunction, denoted $AB$, to which for simplicity we will refer to as the {\it mixed opinion}. The opinion state of the network at the macro level is a vector $\vec{n} = (n_A, n_B, n_{AB}) $, where $n_A, n_B,$ and $n_{AB}$ are the number of nodes holding each of the three opinions. When considering the change in opinion state, $\Delta\vec{n}=(\Delta n_A, \Delta n_B)$, we are only concerned with changes to $n_A$ and $n_B$ since $n_{AB}$ can be found from the constraint $N=n_A+n_B+n_{AB}$.

In the random walk, each interaction is assigned a probability based on the change of the opinion macro state at each time step. A mean field equation can be formed from these probabilities by calculating the expectation of change of macro state which represents the average motion of the system while ignoring any noise arising from randomness. Stability analysis of the system's equilibria may then be performed. 

After developing a drift equation, we shall assume the number of nodes, $N$, in the network is large to removed fluctuations due to noise. The densities of individuals with each opinion state are considered, where $\rho_A=\frac{n_A}{N}$, $ \rho_B=\frac{n_B}{N}$, $\rho_{AB}=\frac{n_{AB}}{N}$ and $1=\rho_A+\rho_B+\rho_{AB}$. 

The models presented in the following subsections are created by expansions of the original naming game with different agent interactions based upon social characteristics of the agents. These variants of the naming game define a two-parameter family with the original naming game being a special case in this family. We also stretch the notion of consensus to allow a {\it consensus} of {\it AB} opinion. What is meant by such consensus is that the equilibrium point is stable and the majority of the nodes hold the mixed opinion {\it AB}.

Following the presentation of the models, linear stability analysis is used to uncover global effects of the variants of the naming game. Prior work in the area of stability in the standard naming game has shown that there exist three equilibrium points: $\{0,1\}$, $\{1,0\}$, and $\{\frac{1}{3}, \frac{1}{3}\}$, where the first two are stable and last is unstable \cite{18, 19}. The equilibrium point $\{\frac{1}{3}, \frac{1}{3}\}$ should not be confused with Baronchelli's threshold of parameter value $\beta = \frac{1}{3}$. These equilibrium points lie in a center manifold where initial values of densities in {\it A} and {\it B} are vacuumed into a line and the system states moves along it to either of the stable points. The consensus the system reaches depends upon where the initial values of densities are located, i.e., whether the initial state lies above or below the line bisecting the first quadrant.

\subsection{Propensity Parameter in the naming game}

\begin{table*}
\footnotesize
\caption{Update events for naming game with parameter $p$ and the associated random walk transition probabilities.}
\begin{tabular}{|p{1.75cm}|p{1.75cm}|p{2cm}|p{1.75cm}|p{9cm}|}
\hline
Speaker & Listener & Event &$\Delta\vec n$ & Probability\\ \hline
B, AB & A & A$\rightarrow$AB & $(-1, 0)$ & $P(A-)=\rho_A (\rho_B+(1-p)\rho_{AB})$\\
A, AB & AB & AB$\rightarrow$A & $(1, 0)$ & $P(A+)=\rho_{AB}(\rho_A+p\rho_{AB})$\\
A, AB & B & B$\rightarrow$AB & $(0,-1)$ & $P(B-)=\rho_B(\rho_A+p\rho_{AB})$\\
B,AB & AB & AB$\rightarrow$B & $(0,1)$ & $P(B+)=\rho_{AB}(\rho_B+(1-p)\rho_{AB})$\\
A,B,AB & A,B,AB & No change & $(0,0)$ & $P(0)=\rho_A(\rho_A+p\rho_{AB})+\rho_B(\rho_B+(1-p)\rho_{AB})$\\ \hline
\end{tabular}
\end{table*}

An interaction in which the speaker holds opinion {\it AB} in the naming game will produce a message with either the opinion {\it A} or {\it B} with equal likelihood. Our first generalization modifies this interaction; a parameter is added which governs the probability of the opinion {\it A} being sent by the speaker. The motivation for introducing this parameter is to enable modeling a network in which some members, lacking a preference to either opinion in the mixed state, will send what they believe is the more accepted opinion if asked to be a speaker. Other ideas may be represented with this parameter, such as influence from media or other means of persuasion of members of the network. 

The propensity parameter, $p$, is the probability that a speaker holding the opinion {\it AB} will send opinion {\it A}. Consequently, $(1-p)$ defines the probability of the speaker sending opinion {\it B}. After each interaction, the listener updates his opinion based on the speaker's message. Each interaction is associated with a particular probability of listener state transition (see Table I). Note that $\rho_{AB}=(1-\rho_A-\rho_B)$.
\\

\begin{figure}[h]
\centering
\begin{subfigure}[b]{0.25\textwidth}
\centering
\includegraphics[width=\textwidth]{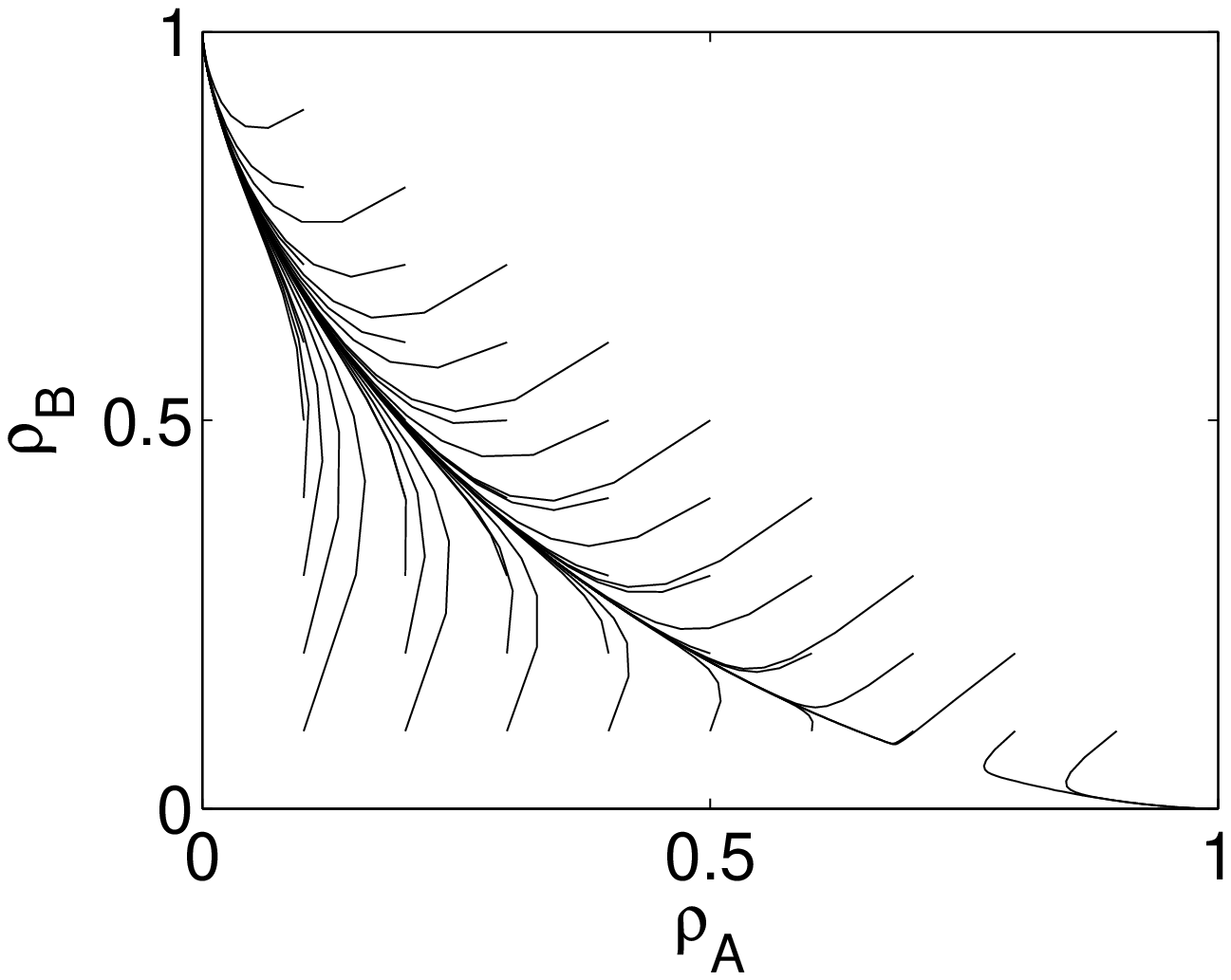}
\caption{$p=0.25$}
\end{subfigure}%
\begin{subfigure}[b]{0.25\textwidth}
\centering
\includegraphics[width=\textwidth]{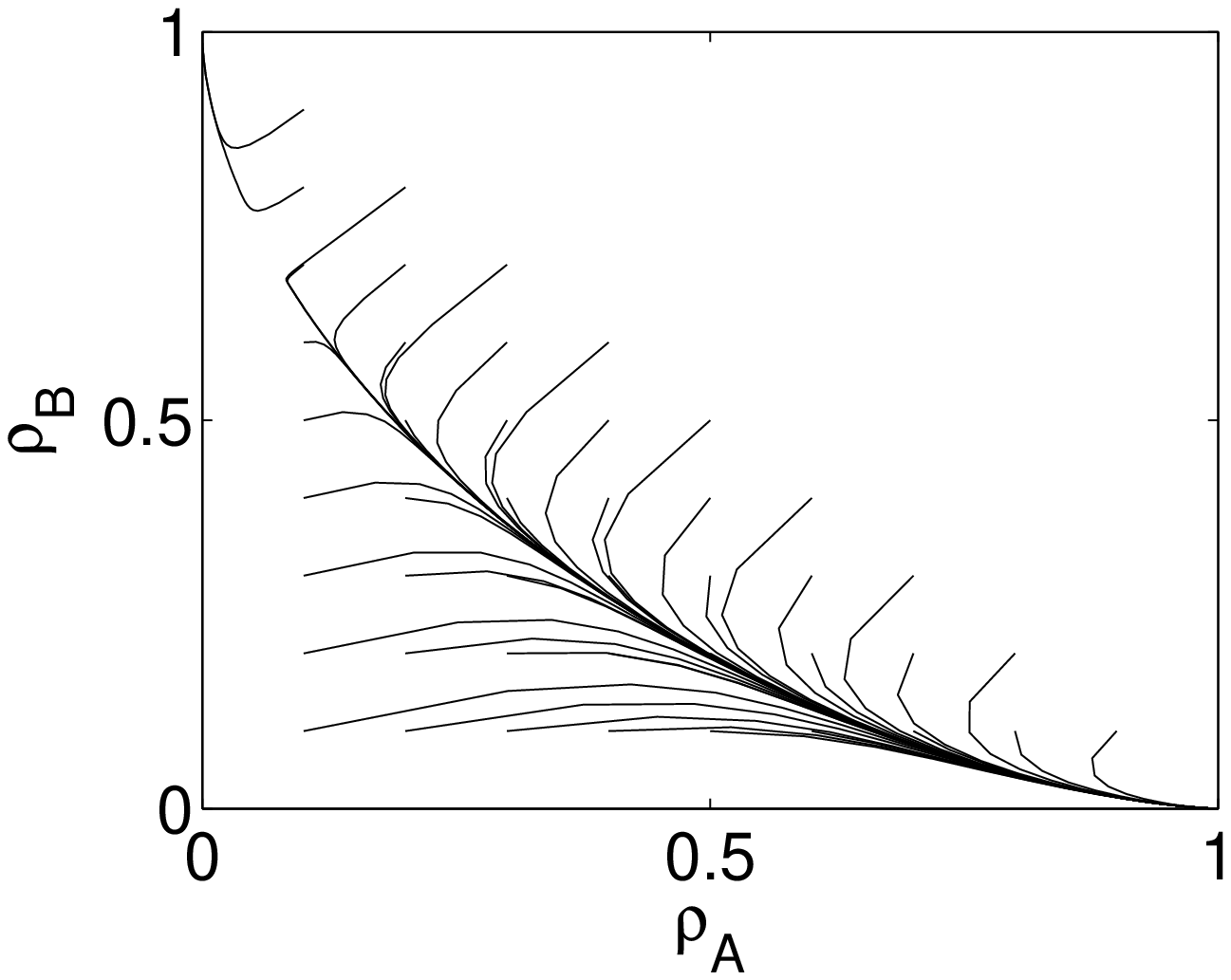}
\caption{$p=0.75$}
\end{subfigure}
\caption{Trajectories of Propensity Models}\label{fig:1}
\end{figure}

Using Table I, we construct a mean-field equation. The system of differential equations defines how the addition of the propensity parameter affects trajectories of the system. The system is described by the following drift equation: 
\begin{equation}
\footnotesize
\begin{aligned}
\frac{d}{dt}
\begin{bmatrix}
\rho_A\\
\rho_B\\
\end{bmatrix}
&=
\begin{bmatrix}
p\rho_{AB}(\rho_A+\rho_{AB})-\rho_A\rho_B\\
\rho_{AB}(\rho_B+(1-p)\rho_{AB})-\rho_B(\rho_A+p\rho_{AB})\\
\end{bmatrix}.
\end{aligned}\end{equation}
Solving for the equilibria of the system, we obtain three equilibrium points: 
$\{0, 1\}$, $\{1, 0\}$, and
$\{\frac{(p-1)^2}{1-p+p^2}, \frac{p^2}{1-p+p^2}\}$. This model shares the equilibrium points with the naming game when $p=\frac{1}{2}$ yielding $\{0,1\}$,$\{1,0\}$, and $\{\frac{1}{3},\frac{1}{3}\}$ as the fixed points.

\begin{figure}[h!]
\includegraphics[width=0.5\textwidth]{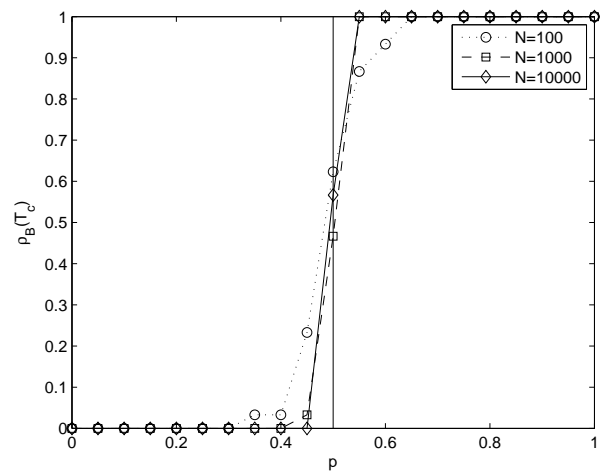}
\caption{Average density of the opinion $B$ at consensus time, $T_c$ as a function of propensity $p$ initialized with a uniform distribution of single opinions with the size of the network $N$}
\end{figure} 

Upon linearization about the fixed points, we obtain the matrix: 
\[\footnotesize\begin{bmatrix}
\rho_B^*(p-1)-p & \rho_A^*(p-1)+2p(\rho_B^*-1)\\
-2(\rho_A^*(p-1)+1)-p(\rho_B^*-2) & p(1-\rho_A^*)-1\\
\end{bmatrix}.\]
The equilibrium points $\{0, 1\}$,$\{1, 0\}$ are stable for all values of $p$, while the third equilibrium point remains a saddle point except when it degenerates at parameter values $p=1$ and $p=0$. This is because as $p$ tends to $1$, or $0$, the third equilibrium point merges with $\{1, 0\}$, or $\{0, 1\}$, respectively. Analytically, the third point follows the parametric curve $\{\frac{(p-1)^2}{1-p+p^2}, \frac{p^2}{1-p+p^2}\}$ which connects the two stable equilibrium points. When $p > 1/2$ the saddle point is closer to $\{1, 0\}$ which causes more $A$ opinion holders to be present, ultimately leading to an {\it A} consensus. Likewise when $p < 1/2$, the saddle point is closer to $\{0, 1\}$, drifting the system towards $B$ consensus. This behavior can be seen in Fig. 1 where consensus on $B$ is more likely when parameter values are $0 \leq p < \frac{1}{2}$ while consensus on $A$ is more likely for parameter values $\frac{1}{2} < p \leq 1$.

Since the propensity parameter $p$ is a measure of which opinion a member holding mixed opinion believes to be more accepted, the drift seen from the stability analysis fits naturally. As the mixed opinion holders follow an inclination that one opinion surpasses the other, with time more and more members will hold the opinion perceived as better. 

An interpretation for $p$ parameter is a polarization of the stable points $\{1, 0\}$,$\{0, 1\}$. As the saddle point moves closer to one of the stable points, it pushes more and more members towards the opposite stable point. When members in mixed state have a higher propensity to say opinion {\it A} when they are speakers, the average opinion such a member shares with listener shifts towards that opinion.

Fig. 2 shows the average over twenty runs of the density of opinion $B$ in the network at consensus state when the network is initialized with a uniform distribution of opinions. Near the parameter value $p=0.5$ a fast transition occurs from consensus on opinion $B$ to consensus on opinion $A$. As the size of the network $N$ increases, the densities of opinion $B$ undergo increasingly fast transition and the plots of these densities as a function of propensity parameter $p$ converge to the Heaviside step function. This implies that propensity controls to which opinion consensus converges by creating with the drift shown in the stability analysis of the mean field approximation of the model dynamics and visible in trajectories of mean field (Fig. 1).

It should be noted that by definition, when $p=\frac{1}{2}$ the propensity model becomes the naming game.

\subsection{Stickiness Parameter in the naming game}

\begin{table*}
\footnotesize
\caption{Update events for the naming game with parameter s and associated random walk transition probabilities.}
\begin{tabular}{|p{1.75cm}|p{1.75cm}|p{2cm}|p{1.75cm}|p{9cm}|}
\hline
Speaker & Listener & Event &$\Delta\vec n$ & Probability\\ \hline
B or AB & A & A$\rightarrow$AB & $(-1, 0)$ & $P(A-) = \rho_A(\rho_B+\frac{1}{2}\rho_{AB})$\\
A or AB & AB & AB$\rightarrow$A & $(1,0)$ & $P(A+)=(1-s)\rho_{AB}(\rho_A+\frac{1}{2}\rho_{AB})$\\ 
A or AB & B & B$\rightarrow$AB & $(0, -1)$ & $P(B-) =\rho_B(\rho_A+\frac{1}{2}\rho_{AB})$\\
B or AB & AB & AB$\rightarrow$B & $(0,1)$ & $P(B+) =(1-s)\rho_{AB}(\rho_B+\frac{1}{2}\rho_{AB})$\\
A,B, or AB & = Speaker & No change & $(0,0)$ & $P(0) =(\rho_A+\frac{1}{2}\rho_{AB})(\rho_A+s\rho_{AB})+(\rho_B+\frac{1}{2}\rho_{AB})(\rho_B+s\rho_{AB})$\\ \hline
\end{tabular}
\end{table*}

It is possible to have individuals in the network who will resist accepting any unique opinion when in a mixed state as they prefer not to side with either one. Following this idea, the corresponding model redefines the interaction with a listener that holds the mixed opinion. In the naming game, when such interaction occurs, the listener changes its state to the speaker's opinion. In the extended model, the listener may resist changing to a single opinion and instead may remain in its mixed state. This extension to the naming game is accomplished with a simple addition of a parameter defining probabilities of two possible outcomes of any interaction involving a listener in the mixed state. This parameter, denoted $s$, can be interpreted as the measure of stickiness to mixed opinion by a listener holding it.  Thus, $s$ defines the probability that the listener will keep its mixed opinion regardless of the interaction with the speaker. Consequently, the value $(1-s)$ defines the probability that a listener will change its opinion to the unique opinion received. The parameter $s$ maps onto parameter $\beta$ defined earlier in \cite{18,19} by a simple transformation $s = (1-\beta)$. 
\begin{figure}[h]
\centering
\begin{subfigure}[b]{0.25\textwidth}
\centering
\includegraphics[width=\textwidth]{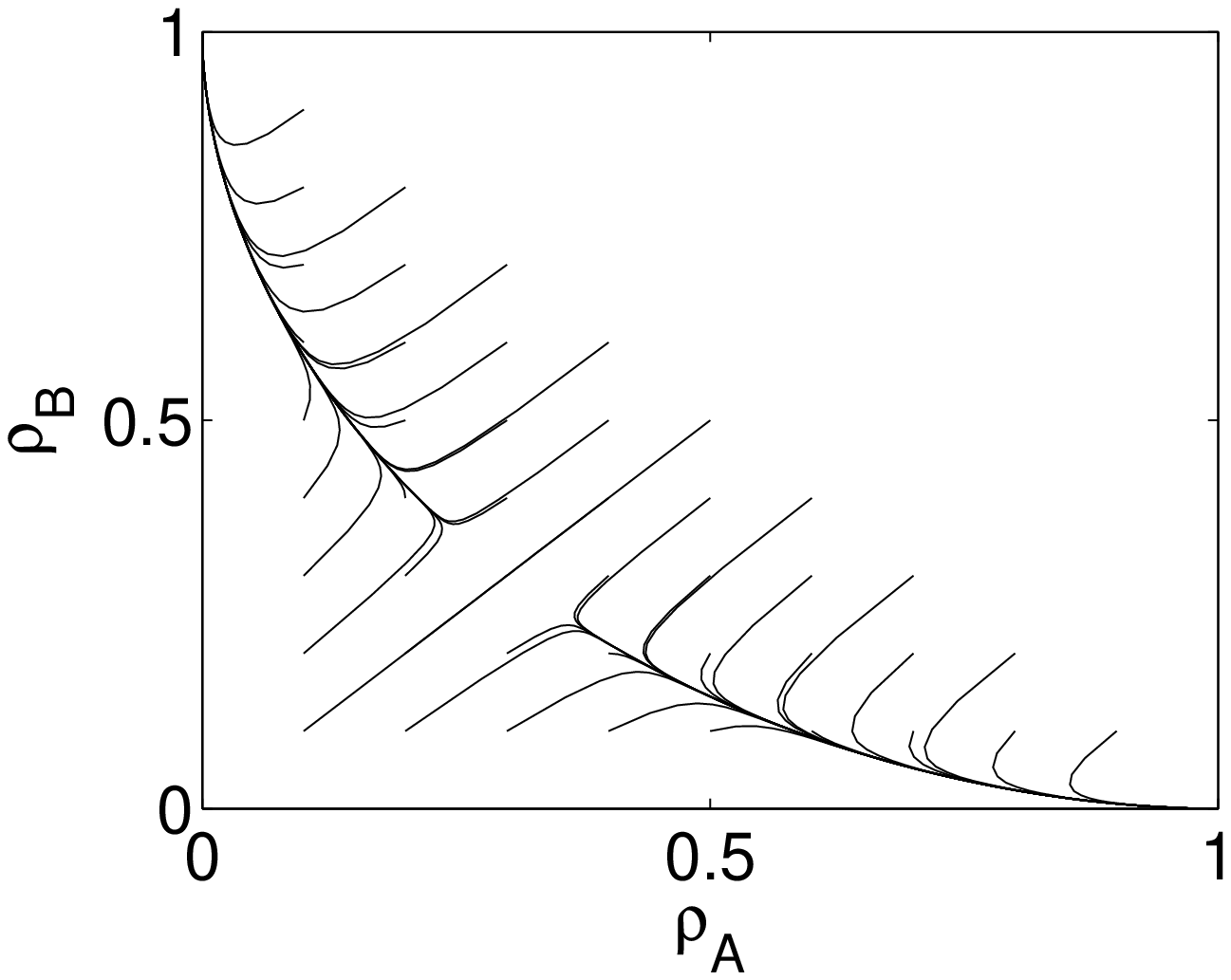}
\caption{$s=0.25$}
\end{subfigure}%
\begin{subfigure}[b]{0.25\textwidth}
\centering
\includegraphics[width=\textwidth]{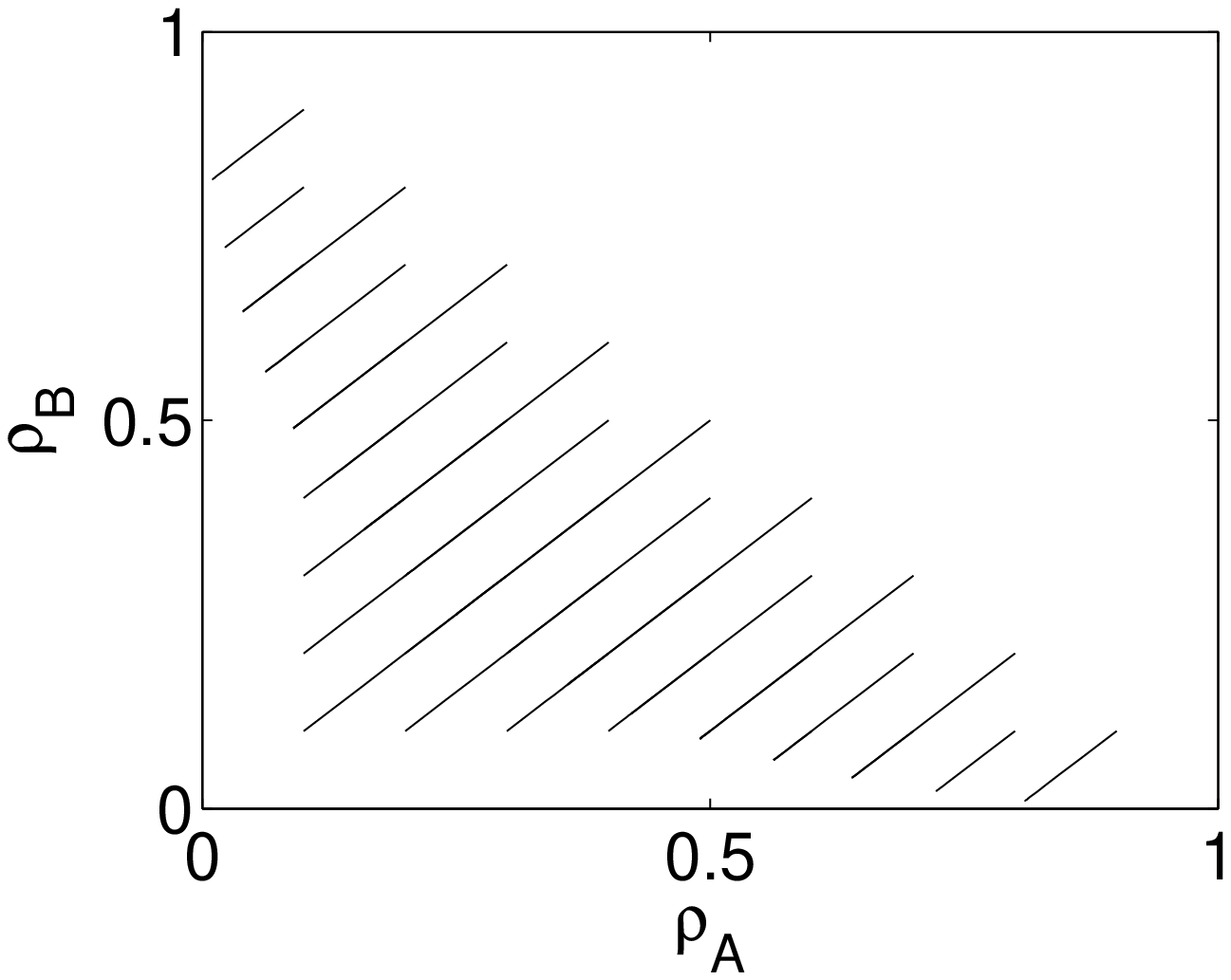}
\caption{$s=.5$}
\end{subfigure}
\begin{subfigure}[b]{0.25\textwidth}
\centering
\includegraphics[width=\textwidth]{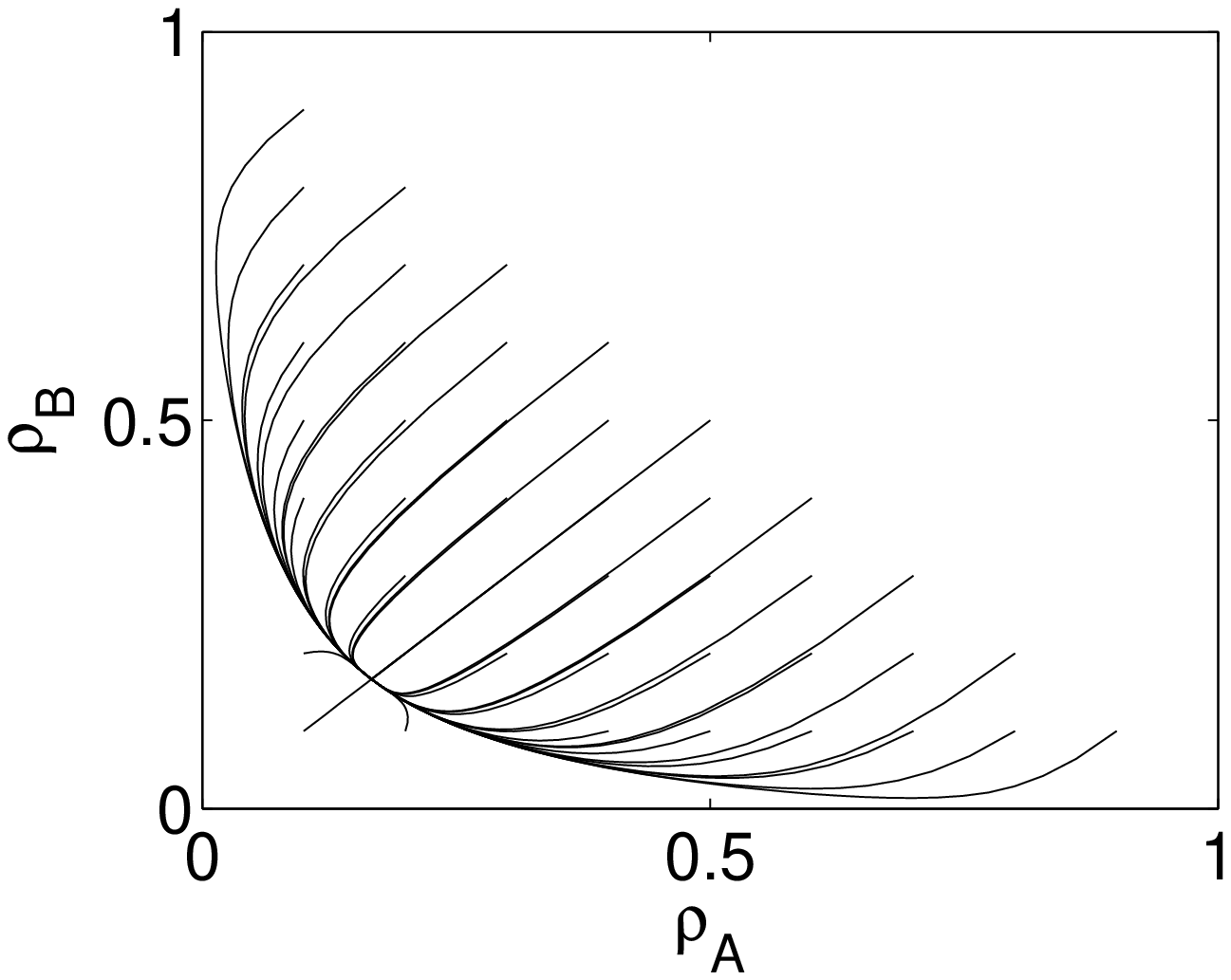}
\caption{$s=0.75$}
\end{subfigure}
\caption{Trajectories of Stickiness Model}\label{fig:2}
\end{figure}

Like in the previous section, we use the probabilities of the listener state transitions (Table II) to construct a mean-field equation: 

\begin{equation}
\footnotesize
\begin{aligned}
\frac{d}{dt}
\begin{bmatrix}
\rho_A\\
\rho_B\\
\end{bmatrix}
&=
\begin{bmatrix}
\frac{1}{2}(1-s)\rho_{AB}(1+\rho_A-\rho_B)-\frac{1}{2}\rho_A(1+\rho_B-\rho_A)\\
\frac{1}{2}(1-s)\rho_{AB}(1+\rho_B-\rho_A)-\frac{1}{2}\rho_B(1+\rho_A-\rho_B)\\
\end{bmatrix}.
\end{aligned}\end{equation}

\begin{figure}[h!]
\includegraphics[width=0.5\textwidth]{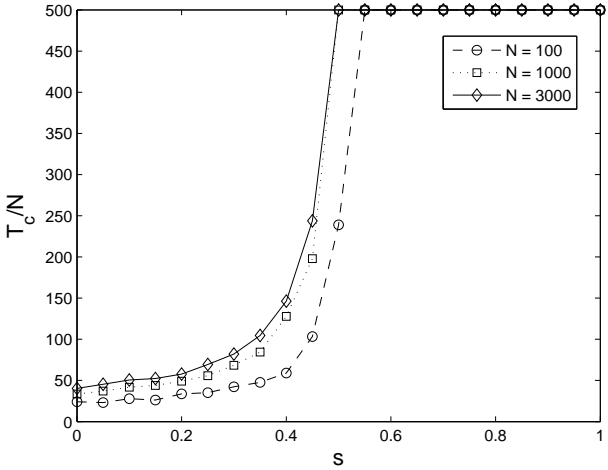}
\caption{Average ratio of time to consensus, $T_c$, to size of the network $N$ as a function of stickiness to mixed opinion $s$ initialized with a uniform distribution of single opinions.}
\end{figure} 

\begin{table*}
\footnotesize
\caption{Update events for naming game with parameters $p$ and $s$, and associated random walk transition probabilities.}
\begin{tabular}{|p{1.75cm}|p{1.75cm}|p{2cm}|p{1.75cm}|p{9cm}|}
\hline
Speaker & Listener & Event &$\Delta\vec n$ & Probability\\ \hline
B or AB & A & A$\rightarrow$AB & $(-1, 0)$ & $P(A-) =\rho_A(\rho_B+(1-p)\rho_{AB})$\\
A or AB & AB & AB$\rightarrow$A & $(1, 0)$ & $P(A+)=(1-s)\rho_{AB}(\rho_A+p\rho_{AB})$\\
A or AB & B & B$\rightarrow$AB & $(0, -1)$ & $P(B-) = \rho_B(\rho_A+p\rho-{AB})$\\
B or AB & AB & AB$\rightarrow$B & $(0, 1)$ & $P(B+) =(1-s)\rho_{AB}(\rho_B+(1-p)\rho_{AB})$\\
A,B, or AB & = Speaker & No change & $(0,0)$ & $P(0) =(\rho_A+p\rho_{AB})(\rho_A+s\rho_{AB})+(\rho_B+(1-p)\rho_{AB})(\rho_B+s\rho_{AB})$\\ \hline
\end{tabular}
\end{table*}

Its equilibria are $\{0, 1\}$,$\{1, 0\}$, and 
$\{\frac{s-1}{2s-3}, \frac{s-1}{2s-3}\}$.
By definition, when $s=0$, the model becomes the naming game whose well-known equilibria are $\{0,1\}$,$\{1,0\}$, and $\{\frac{1}{3},\frac{1}{3}\}$. After linearizing the system, we obtain:
\[\footnotesize\begin{bmatrix}
-\frac{1}{2}+s\rho_A^*-\frac{1}{2}\rho_B^* & -\frac{1}{2}\rho_A^*+(s-1)(1-\rho_B^*)\\
(s-1)(1-\rho_A^*)-\frac{1}{2}\rho_B^* & -\frac{1}{2}-\frac{1}{2}\rho_A^*+s\rho_B^*\\
\end{bmatrix}.\]

Unlike in the propensity model, where the saddle point traveled from consensus on {\it A} to consensus on {\it B} for values of $p$ varying from $0$ to $1$, the stability of the points will change from two stable points and a saddle point into one stable point and two saddle points. The new stable node will also travels towards the origin. These two sets of stabilities can be observed in two intervals of parameter $s$ that are: $[0, \frac{1}{2})$ and $(\frac{1}{2}, 1]$ as portrayed in Fig. 3.

Inspecting the interval $[0,\frac{1}{2})$ of parameter s values, we notice that the points $\{0,1\}$ and $\{1,0\}$ are stable while $\{\frac{s-1}{2s-3},\frac{s-1}{2s-3}\}$ is a saddle. The trajectories for these values of parameter $s$ are similar to that of the NG model except that as $s$ moves towards $\frac{1}{2}$, the center manifold of the system bends closer towards the origins. 

In the interval $(\frac{1}{2},1]$ of parameter $s$ values, the saddle point $\{\frac{s-1}{2s-3},\frac{s-1}{2s-3}\}$ becomes a stable point and the two stable points $\{1,0\}$ and $\{0,1\}$ become saddle points. 
Thus, when the stickiness increase to $s=1$, all the mixed opinion nodes stop listening to outside information and remain mixed. The whole system eventually converges to the mixed opinion. 

In this interval (i.e. for $\frac{1}{2}<s\leq 1$), no longer the consensus on $A$ or $B$ happens because, thanks to stickiness, there are always some nodes 
in mixed opinion state present. In the ``Original'' NG model, as established in \cite{18}, consensus is not possible for $0<\beta<\frac{1}{3}$ where $\beta = 1-s$, so for $\frac{2}{3} < s < 1$. This is consistent with the difference in models used. The interactions between pairs of nodes in mixed state remove nodes in mixed opinion state twice as fast as in ``Listener-only'' NG model used here, Hence, keeping nodes in mixed opinion state surviving, the ``Original'' NG model requires higher values of $s >\frac{2}{3}$ than  $s > \frac{1}{2}$ needed for NG model used here. The consistent presence of nodes in the mixed state in this interval motivated us to generalizing the notion of a consensus. Instead of a consensus in which all nodes hold the same opinion, a steady state arises in which the majority of the nodes hold the mixed opinion while, on average, a fixed number of unique opinions $A$ and $B$ holders remain.  

To complete the analysis, when parameter $s=\frac{1}{2}$ the equilibrium point $\{\frac{s-1}{2s-3}, \frac{s-1}{2s-3}\}$ becomes degenerate. However, through observation of the approximate trajectories (Fig. 3), we conclude there is no attraction to or repulsion from the points $\{0, 1\}$ and $\{1, 0\}$. Instead all trajectories move towards a hyperbolic curve. The parameter value $s=\frac{1}{2}$ is a pitchfork bifurcation point which can be shown with an appropriate change of variables. 

Figure 4 shows the normalized time to consensus $T_c$/$N$ taken over twenty runs. Values of $T_c$/$N = 500$ are in fact lower bounds of the normalized consensus time since this value was used to stop the simulations if the consensus was not reached by then. As the parameter $s$ increases, a fast transition occurs after which the network no longer can reach consensus. As the size of the network $N$ increases, the location of this fast transition recedes towards $s=0.5$ as suggested by the inversion of stabilities in the mean field valid when there exists a large population of mixed opinion holders. These results also suggest that the effect of stickiness is stronger for large networks where smaller values of the parameter $s$ will result in a loss of consensus while smaller networks require stronger stickiness for the inversion seen in the mean field to occur. 

By definition, at $s=0$, the system reverts to the NG model where $\{0,1\}$ and $\{1,0\}$ are stable points and $\{\frac{1}{3},\frac{1}{3}\}$ is a saddle point.
\begin{table*}
\footnotesize
\caption{Update events for naming game with parameter $s$ and associated random walk transition probabilities.}
\begin{tabular}{|p{1.75cm}|p{1.75cm}|p{2cm}|p{1.75cm}|p{9cm}|}
\hline
Speaker & Listener & Event &$\Delta\vec n$ & Probability\\ \hline
B or AB & A & A$\rightarrow$AB & $(-1, 0)$ & $P(A-) = \rho_A(\rho_B+\frac{1}{2}\rho_{AB})$\\
A or AB & AB & AB$\rightarrow$A & $(1, 0)$ & $P(A+)=(1-s)\rho_{AB}(\rho_A+c_A+\frac{1}{2}\rho_{AB})$\\
A or AB & B & B$\rightarrow$AB & $(0, -1)$ & $P(B-) =\rho_B(\rho_A+c_A+\frac{1}{2}\rho_{AB})$\\
B or AB & AB & AB$\rightarrow$B & $(0, 1)$ & $P(B+) =(1-s)\rho_{AB}(\rho_B+\frac{1}{2}\rho_{AB})$\\
A,B, or AB & = Speaker & No change & $(0, 0)$ & $P(0) =(\rho_A+c_A+\frac{1}{2}\rho_{AB})(\rho_A+c_A+s\rho_{AB})+(\rho_B+\frac{1}{2}\rho_{AB})(\rho_B+s\rho_{AB})$\\ \hline
\end{tabular}
\end{table*}

\subsection{Propensity and Stickiness Parameters in the naming game}

The socially motivated interactions presented in the two previous sections may co-exist within the same network. Hence, here we present a model that incorporates both propensity and stickiness. Each parameter is identical to its counterpart introduced in one of the two previous sections. The people holding mixed opinion may be influenced by some outside source, as reflected by parameter $p$, while still maintaining the stickiness, $s$, to mixed opinion. It should be noted that for some values of parameters $s$ and $p$, the positions of equilibria are outside of the feasible range of densities of nodes in different states. These values are discussed briefly in this section. 
\begin{figure}[h]\footnotesize
\centering

\begin{subfigure}[t]{0.25\textwidth}
\centering
\includegraphics[width=\textwidth]{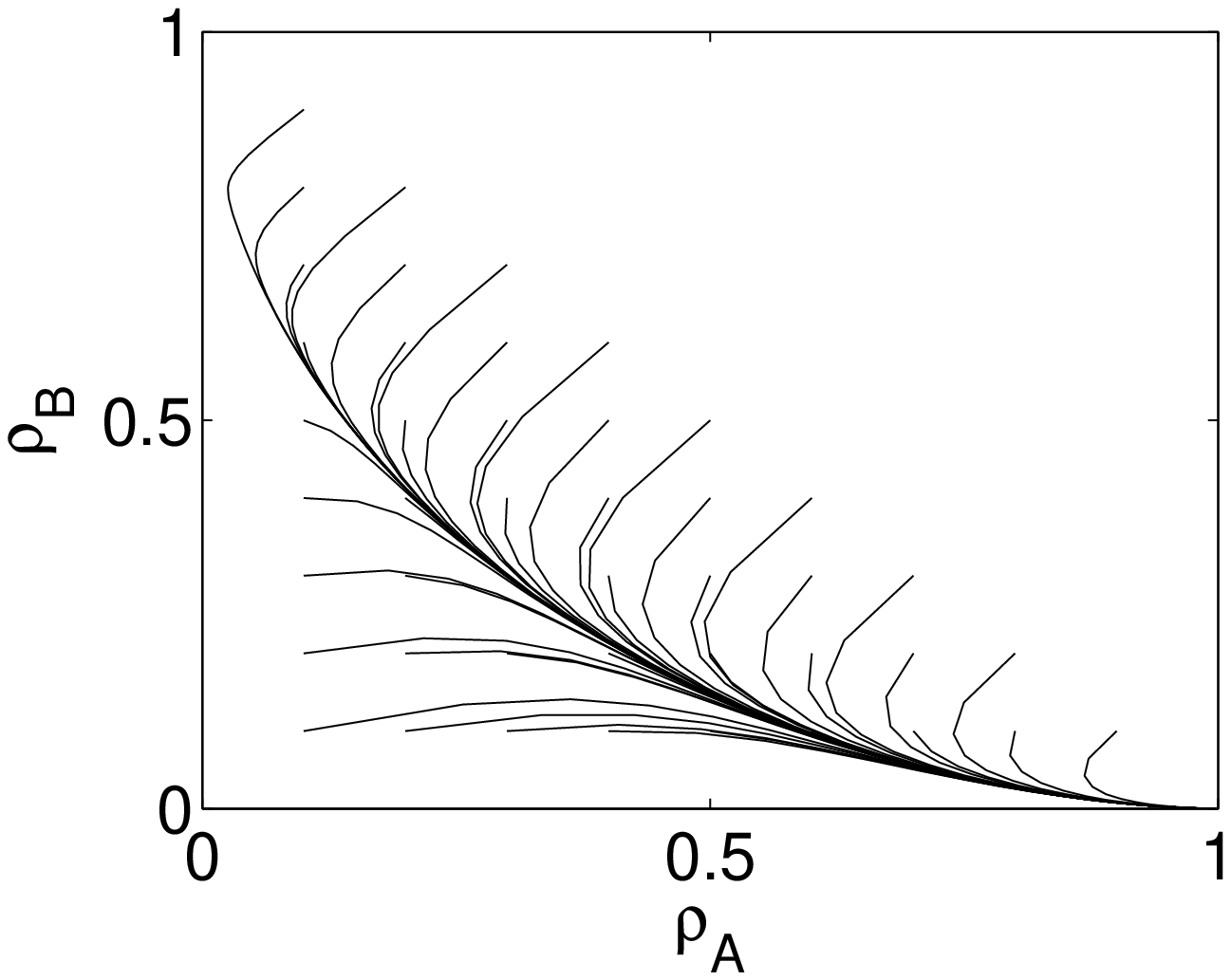}
\caption{s=0.25, p=0.75}
\end{subfigure}%
\begin{subfigure}[t]{0.25\textwidth}
\centering
\includegraphics[width=\textwidth]{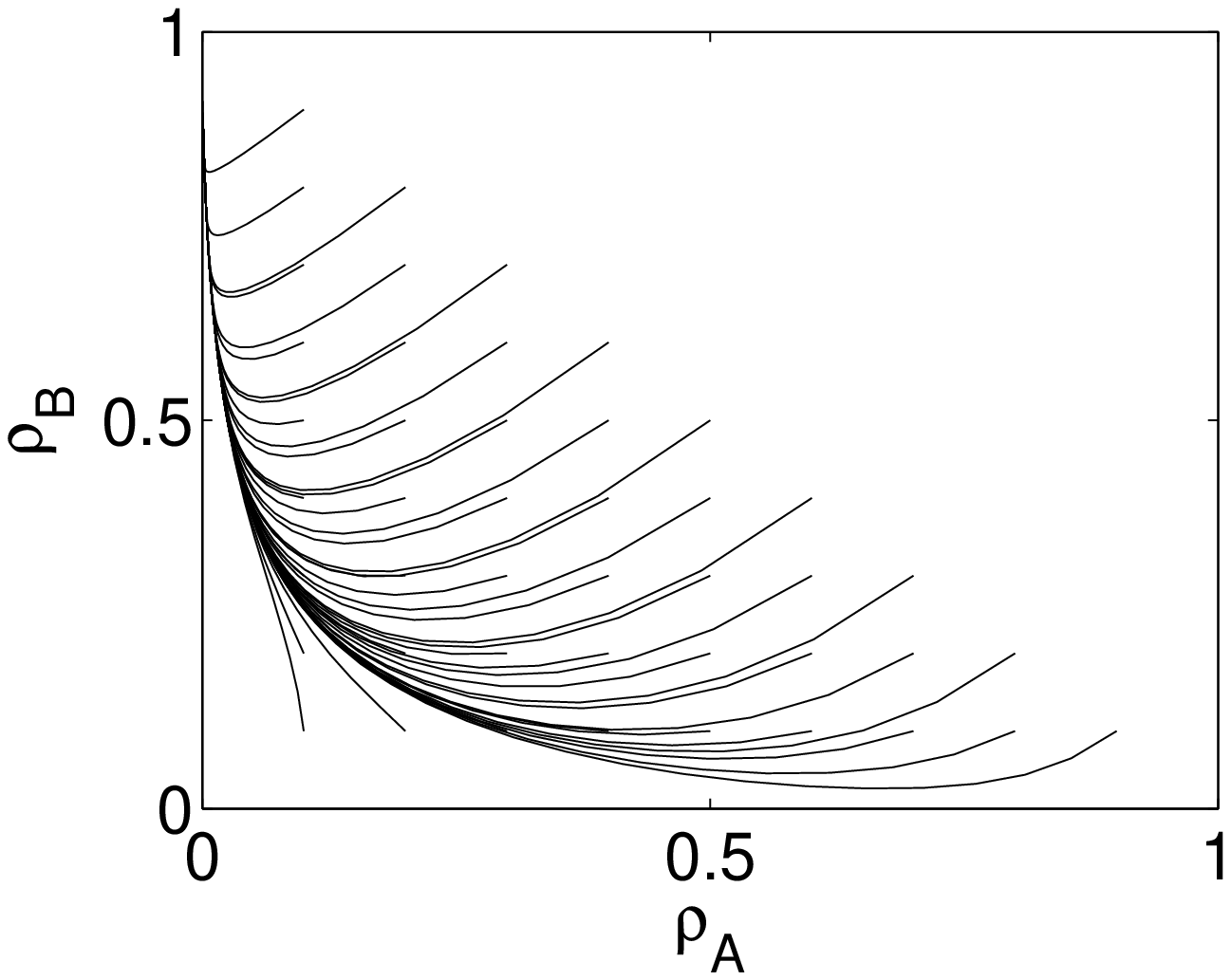}
\caption{s=0.75, p=0.25}
\end{subfigure}
\begin{subfigure}[t]{0.24\textwidth}
\centering
\includegraphics[width=\textwidth]{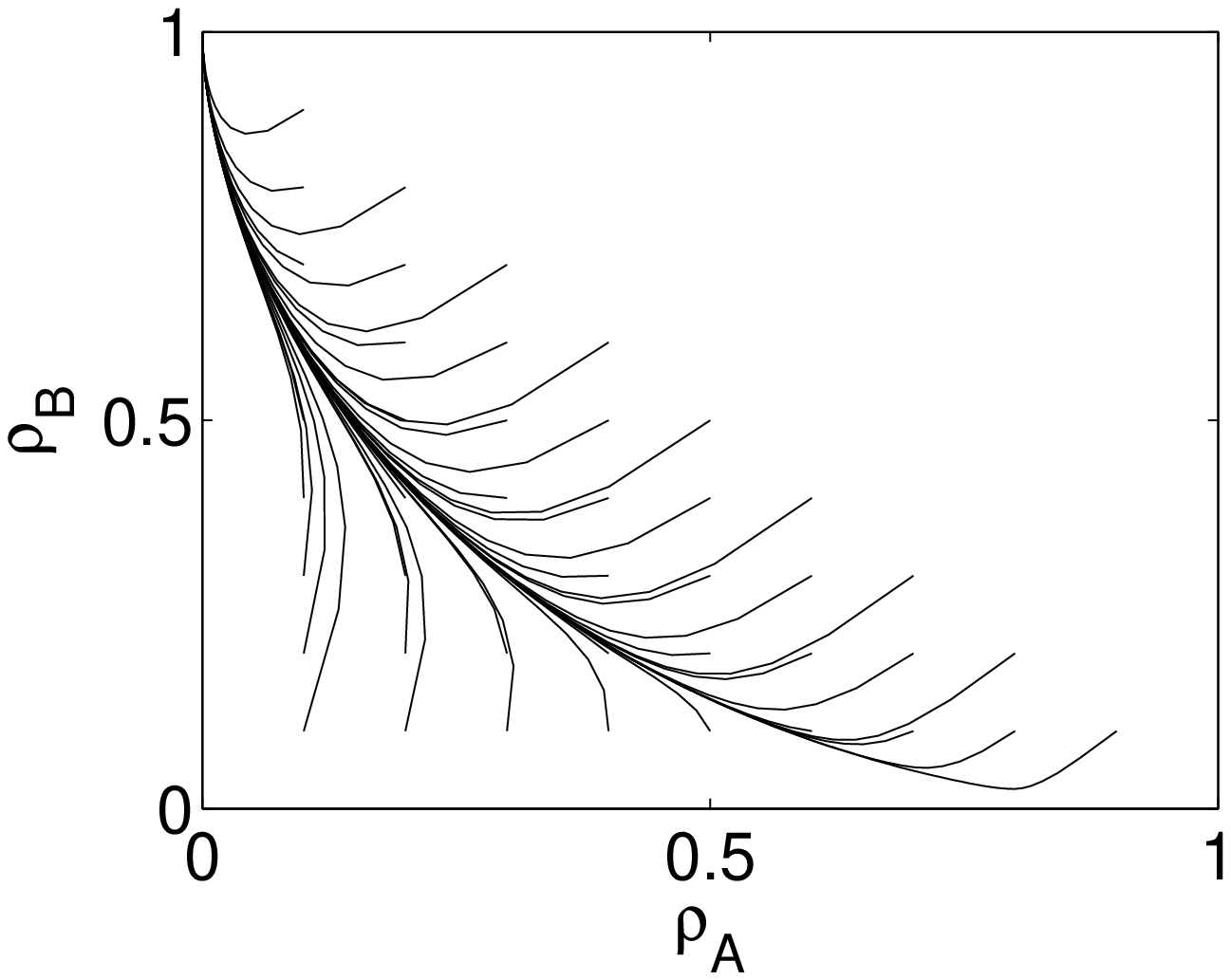}
\caption{s=0.25, p=0.25}
\end{subfigure}%
\begin{subfigure}[t]{0.24\textwidth}
\centering
\includegraphics[width=\textwidth]{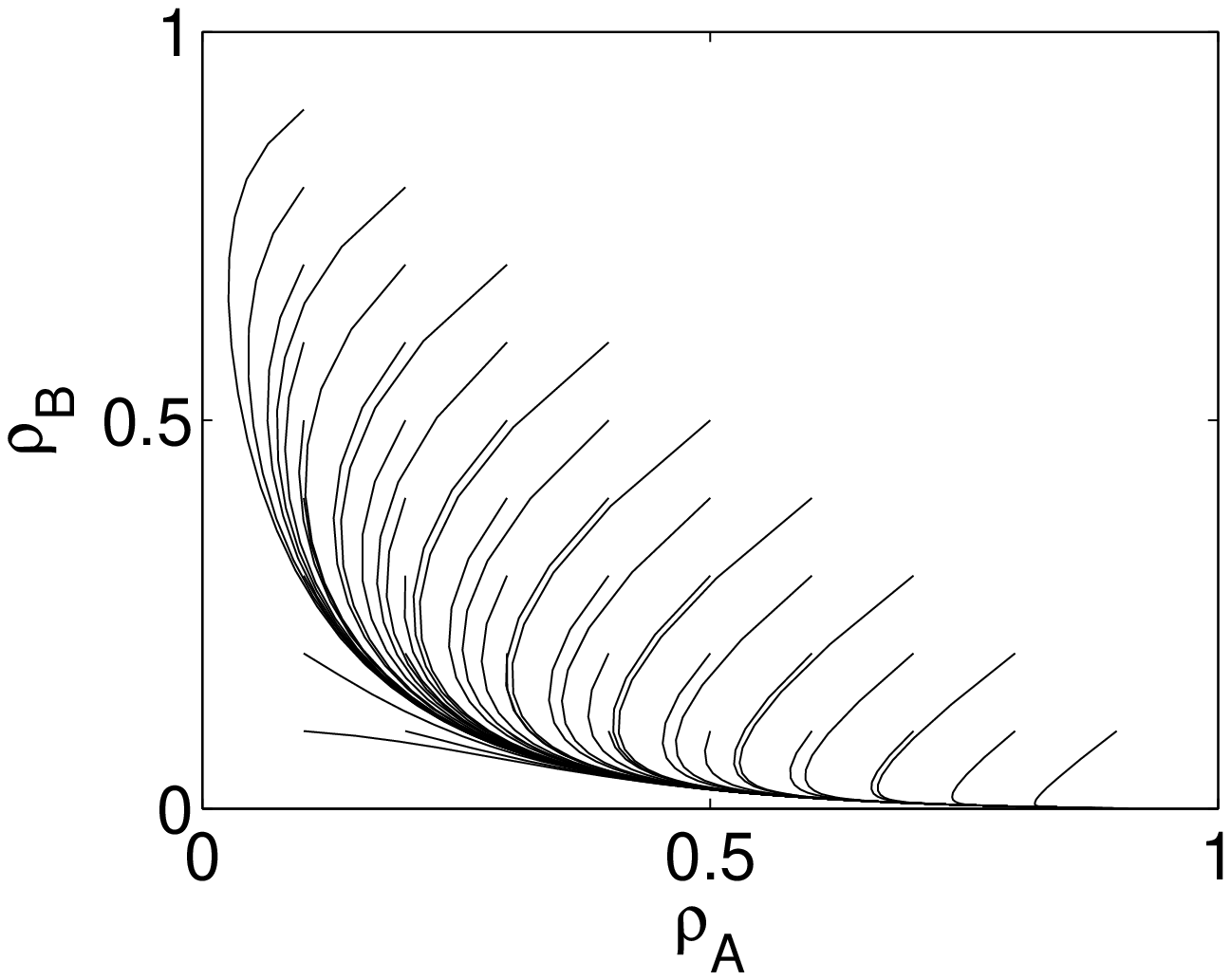}
\caption{s=0.75, p=0.75}
\end{subfigure}
\caption{Trajectories of Two-Parameter naming game}\label{fig:3}
\end{figure}

The mean-field for this variation of the NG model (using Table III) is: 
\begin{equation}
\footnotesize
\begin{aligned}
\frac{d}{dt}
\begin{bmatrix}
\rho_A\\
\rho_B\\
\end{bmatrix}
&=
\begin{bmatrix}
(1-s)\rho_{AB}(\rho_A+p\rho_{AB})-\rho_A(\rho_B+(1-p)\rho_{AB})\\
(1-s)\rho_{AB}(\rho_B+(1-p)\rho_{AB})-\rho_B(\rho_A+p\rho_{AB})\\
\end{bmatrix}.
\end{aligned}\end{equation}

Solving for the equilibria of the propensity and stickiness model yields the points $\{0,1\}$,$\{1,0\}$, and $\{\frac{(s-1)(p+s-1)^2}{(2s-1)(p^2-p+(s-1)^2)},\frac{(s-1)(p-s)^2}{(2s-1)(p^2-p+(s-1)^2)}\}$. By definition, when $s=0$, $p=1/2$ the system becomes the NG model with equilibria points $\{0,1\}$, $\{1,0\}$, and $\{\frac{1}{3},\frac{1}{3}\}$.

In this model, there are two stationary equilibria which do not vary with different values of parameters $s$ and $p$. However, the third point presents a few drawbacks to the model. Half of the values of parameters p and s yield unfeasible densities of nodes in different states; those are densities that do not satisfy constraints such as $1=\rho_A+\rho_B+\rho_{AB}$. Since the focus of this paper is on linear stability, we consider only such combinations of values of parameters $s$ and $p$ that satisfy the node density range and stability criteria. With this clarification, there are two regions in the $p-s$ plane satisfying the criteria: region (1) $ (1-s)>p$ intersected with $ p> s$, and region (2): $p> (1-s)$ intersected with $p<s$.

For the values within the hourglass-like region representing the union of both regions, the stability of the third equilibrium point combines behavior of both the propensity and stickiness models (Fig. 5). The dominating variable in the combined model is the parameters $s$ of the stickiness model. While propensity affects how the equilibrium point swings from one consensus point to another, the stickiness directly affects the stability of the equilibria.

There are two regions of equilibrium points, one with $s$ in the interval $[0, 1/2)$ and another with $s$ in the complementary interval $(1/2,1]$. In the first interval, the first two equilibria are stable while the third is a saddle point. However in the second interval, we conclude, as we did in section II.B, that for values of $s$ in $(1/2, 1]$ the third equilibrium point becomes stable while the first two become saddle points while the consensus points $(0, 1)$ and $(0, 1)$ remain saddle points. This change of equilibria occurs when $s=1/2$. 

\begin{figure}[h]
\centering
\includegraphics[width=0.5\textwidth]{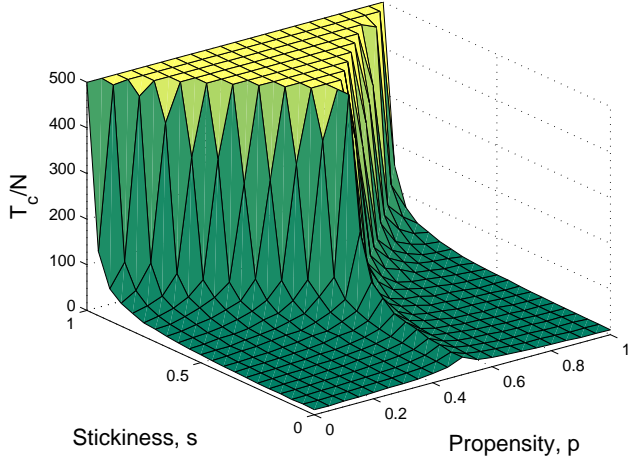}
\caption{Average ratio of consensus time,$T_c$, to the network size, $N$ as a surface of propensity, $p$, and stickiness, $s$. Lighter color implies longer consensus time.}
\end{figure}

\begin{figure}[h]
\centering
\includegraphics[width=0.5\textwidth]{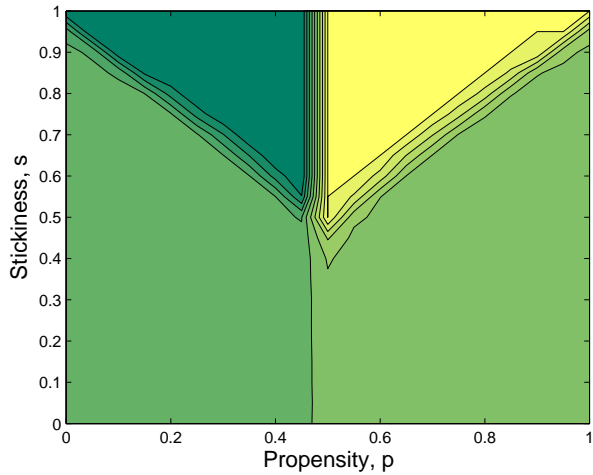}
\caption{Weighted average ratio of consensus time,$T_c$, to the network size, $N$ as a contour of propensity, $p$, and stickiness, $s$. Darker color implies consensus towards $B$ and lighter color implies consensus towards $A$.}
\end{figure}

Figure 6 and 7 show the average ratio of consensus time, $T_c$, to size of the network, $N$ over twenty runs with the network size of $N = 1000$.  In figure 6, for values of stickiness $s > \frac{1}{2}$ a sharp transition occurs where consensus at opinions $A$ and $B$ is changed with a majority of $AB$ holders as seen in the previously discussed region (2). In figure 7,  a weight was added to distinguish which consensus, $A$ or $B$, occurs. For values of propensity $p < \frac{1}{2}$, consensus occurs most often with opinion $B$. Likewise, consensus occurs at the opinion $A$ for values of propensity $p > \frac{1}{2}$. 

\section{Committed Agents in the Stickiness Model}

Committed agents \cite{30, 31} are holders of a unique opinion that they preserve regardless of interactions in which they are listeners. Yet, as speakers, they send their opinion and can influence holders of other opinions. Understanding how committed minorities in the naming game affect the consensus is a fundamental issue of NG dynamics. The notion of committed agents has also received attention in the Voter Model \cite{32}. The important aspects of the study of committed agents include the time to consensus and the minimum density of committed agents necessary for fast consensus \cite{33}. Naturally, an addition of committed agents imposes a strong drift to the opinion that these agents hold. The general rules for constructing the drift equation do not differ from those of the normal model except when the chosen listener is a member of the committed minority. 

\subsection{Stickiness Model with Committed Agents}

The motivation behind adding some fraction of committed agents to the stickiness model is to see whether it is possible to cause a {\it consensus on mixed opinion} even in presence of a drift from the introduced committed agents when the stickiness strength grows beyond the bifurcation point discussed earlier. Committed agents will be represented by their density $c_A$ ranging from $0$ to $1$. This addition requires a modification to the density equation, as now we have $1= \rho_A+c_A+\rho_B+\rho_{AB}$. Using the variables defined in Table IV, the mean-field is:

\begin{equation}
\footnotesize
\begin{aligned}
\frac{d}{dt}
\begin{bmatrix}
\rho_A\\
\rho_B\\
\end{bmatrix}
&=
\begin{bmatrix}
(1-s)\rho_{AB}(\rho_A+c_A+\frac{1}{2}\rho_{AB}) -\rho_A (\rho_B+\frac{1}{2}\rho_{AB})\\
(1-s)\rho_{AB}(\rho_B+\frac{1}{2}\rho_{AB})-\rho_B(\rho_A+c_A+\frac{1}{2}\rho_{AB})\\
\end{bmatrix}.
\end{aligned}\end{equation}

\subsection{Stability of Stickiness Model with Committed Agents}
Solving the equation $\frac{d\vec\rho}{dt}=0$ yields the equilibrium points 
$\{1-c_A,0\}$ and the conjugate pair 

\[\scriptsize\begin{Bmatrix}
\frac{(s-1)(2s-1+c_A(7-6s)\mp\sqrt{2s-1}\sqrt{2s-1+c_A^2(2s-1)+2c_A(7-6s)}}{6-16s+8s^2},\\
\frac{(4-11s+6s^2+c_A(-4+5s-2s^2)\mp(2-s)\sqrt{(2s-1)(2s-1+c_A^2(2s-1)-2c_A(6s-7))}}{6-16s+8s^2}.\\
\end{Bmatrix}
\]
Before linearizing the stickiness model and applying stability analysis, the conjugate pair should be inspected. The number of equilibrium points can be reduced to two by setting the discriminant to zero. This imposes an equation \[ c_A=\frac{6s-7\pm4\sqrt{3-5s+2s^2}}{2s-1}\] giving two values for $c_A$. This curve in $c_A-s$ space plane becomes infeasible when one or both $c_A$ values becomes negative as well as when $c_A \geq 1$. For the parameter $s \geq \frac{1}{2}$ both values of $c_A$ are negative which will be discussed later in this section. When the parameter $s$ is $0 \leq s < \frac{1}{2}$, the larger value of $c_A$ is greater than 1 and is infeasible under our constraints. This leaves the smaller value of $c_A$ to which we will refer as $c_A^c$ which satisfies our constraints, defined as
\[ c_{A}^c=\frac{6s-7+4\sqrt{3-5s+2s^2}}{2s-1}.\] The pairs $\{c_A, s\}$ represent some threshold of a bifurcation and $c_{A}^c$ defined above represents some critical fraction of committed agents. 

Linearizing the stickiness with committed agent model about an arbitrary equilibrium point $(\rho_A^*, \rho_B^*)$ yields the following matrix:
\[\scriptsize\begin{bmatrix}
\frac{1}{2}(2s\rho_A^*-\rho_B^*+c_A(2s-1)-1 & -\frac{\rho_A^*}{2}+(1-s)\rho_B^*+s-1\\
(1-s)\rho_A^*-\frac{\rho_B^*}{2}+(1-s)c_A+s-1 & \frac{1}{2}(-\rho_A^*+2s\rho_B^*-(c_A+1))\\
\end{bmatrix}.\]
Using this equation for the threshold, we confirm that indeed the equation gives a critical fraction of committed agents necessary for forcing a steady-state. For an arbitrary $s<\frac{1}{2}$, a saddle-point bifurcation occurs at $c_{A}^c$. When $c_A$ is smaller than $c_{A}^c$ a stable point and a saddle point appear along with the already existing stable point at $\{1-c_A, 0\}$ (Fig. 8(a)). 
\begin{figure}[h]\footnotesize
\centering
\begin{subfigure}[b]{0.25\textwidth}
\centering
\includegraphics[width=\textwidth]{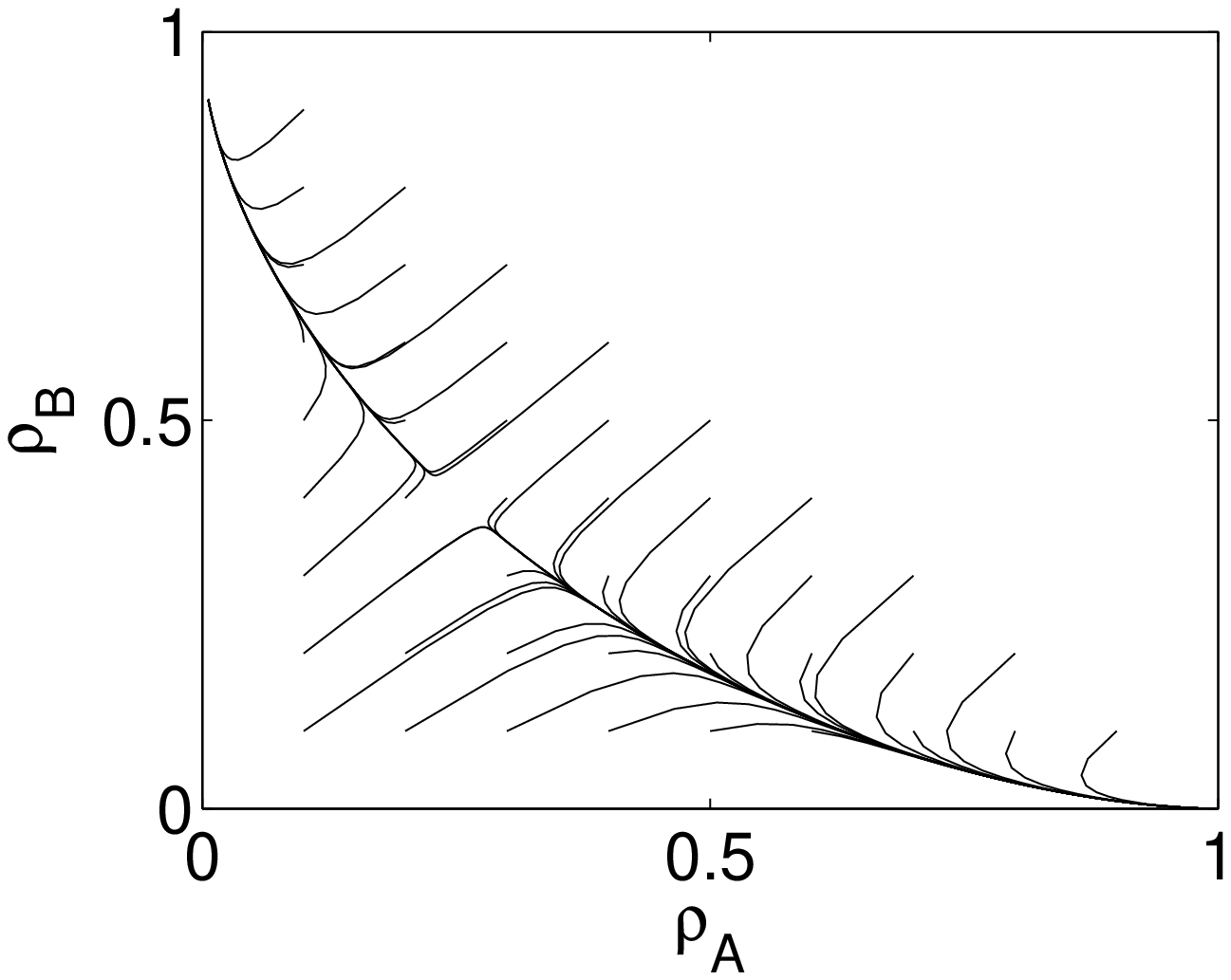}
\caption{s=0.1, $c_A$=0.03}
\end{subfigure}%
\begin{subfigure}[b]{0.25\textwidth}
\centering
\includegraphics[width=\textwidth]{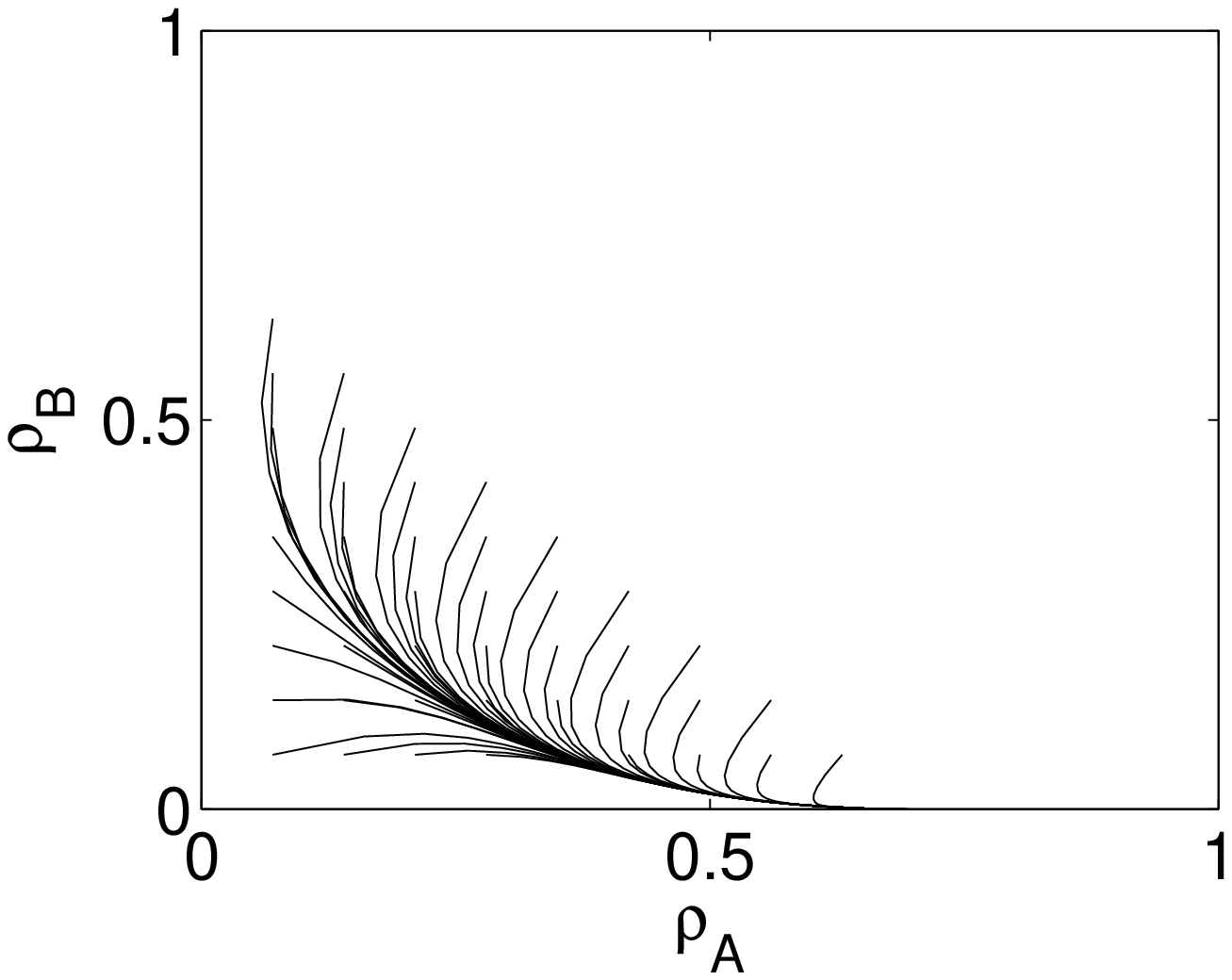}
\caption{s=0.25, $c_A$=0.3}
\end{subfigure}
\begin{subfigure}[b]{0.25\textwidth}
\centering
\includegraphics[width=\textwidth]{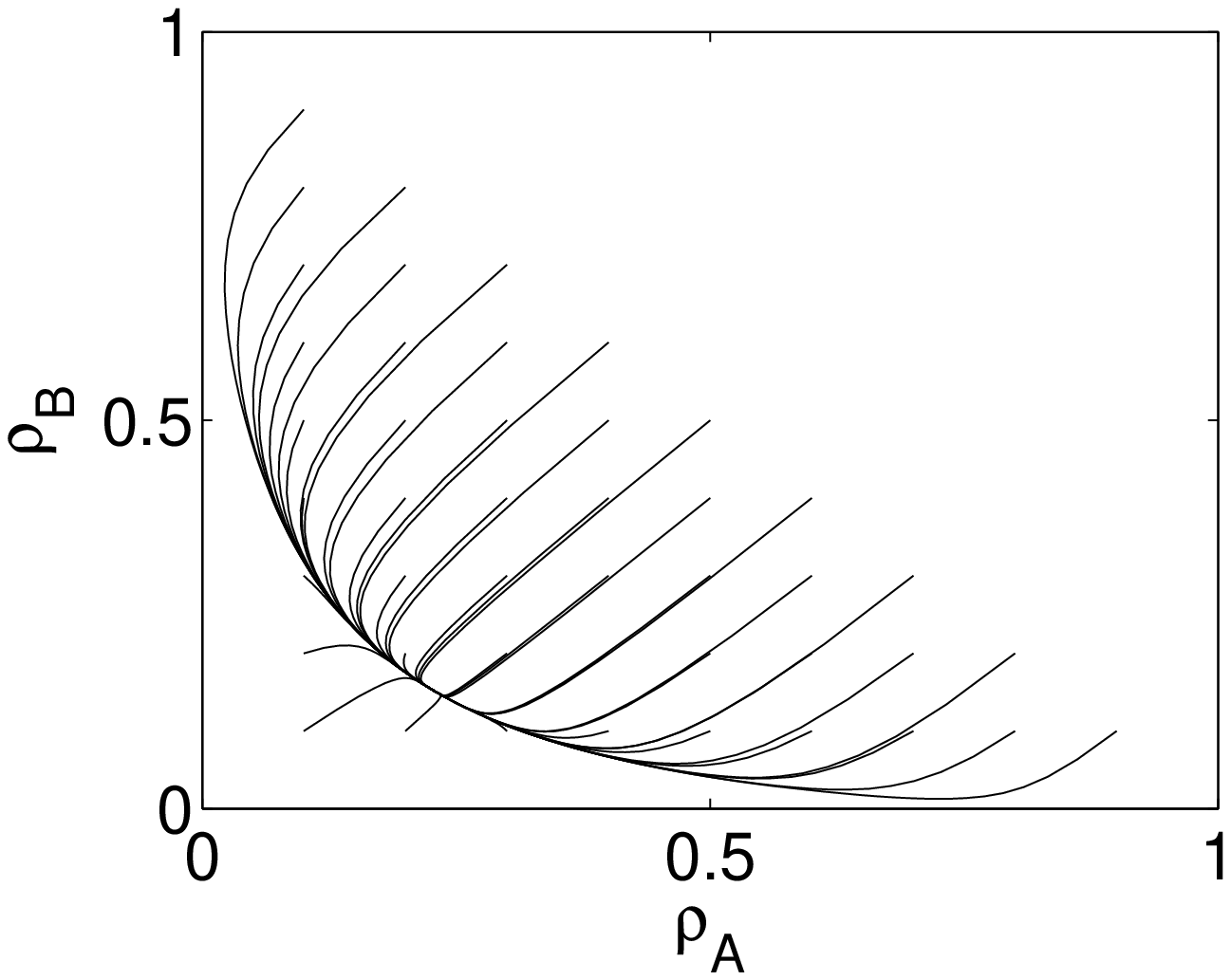}
\caption{s=0.7, $c_A$=0.03}
\end{subfigure}
\caption{Trajectories of Stickiness Model with Committed Agents computed using Runge Kutta}\label{fig:5}
\end{figure}

When $c_A$ is greater than $c_{A}^c$, degeneracies occur in both the location and the stability of the equilibrium point. Though infeasible, a conjugate pair of complex equilibrium points appears whose eigenvalues are also complex numbers. However, the two equilibrium points are unstable spirals whose respective complex components seem to cancel out yielding a saddle node. The numerically plotted trajectories (Fig. 8(b)) support the hypothesis of canceling complex terms and enables us to observe that the system is in an active steady-state while making a flow towards the consensus point $\{1-c_A,0\}$. These results can be seen in Fig. 8(b), however this odd stability is not yet fully analyzed, so we plan to study it further in future work. 

As mentioned earlier, the necessary values of $c_A$ dependent on the parameter $s$ did not satisfy our constraints when the parameter $s > \frac{1}{2}$. The stickiness model without committed agents undergoes a bifurcation at the parameter $s=\frac{1}{2}$ when the saddle point becomes stable and remains so when $s$ increases towards $1$. The same behavior is observed in the stickiness model with committed agents. Since the $A$ opinion consensus in the stickiness model changes it stability from a stable point to a saddle point, the corresponding curve in $c_A-s$ plane can be found by checking the stability of the point $\{1-c_A, 0\}$. When the following eigenvalue \[\frac{1}{2}(-c_A+2s-1),\] which has an eigenvector along the center manifold, changes from negative to positive, a saddle point arises. Setting this eigenvalue to zero and solving for $c_A$ yield the line $c_A=2s-1$. This equation also yields another critical fraction of agents who are able to overcome the pull of stickiness and cause an active steady-state to occur once again. 

\begin{figure}[h]
\centering
\includegraphics[width=0.5\textwidth]{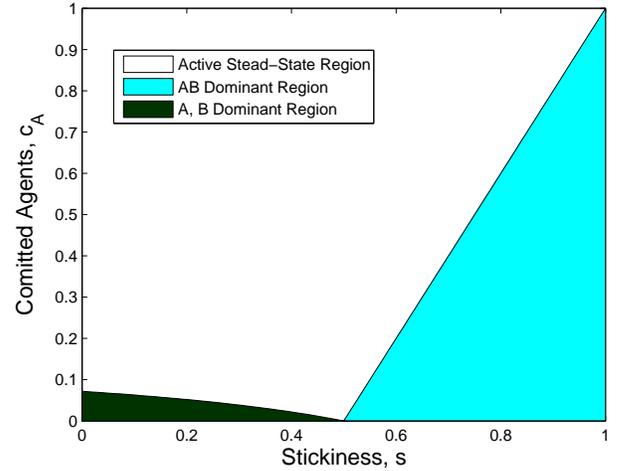}
\caption{Global Stability in $c_A-s$ plane}
\end{figure}

This line is a boundary of stabilities as seen in the case when $c_A=0$. When $c_A$ lies above the boundary, the system resembles the standard naming game model with the exception of the case when the number of committed agents is above the critical point. When $c_A$ is below the boundary, we get a {\it consensus} on the mixed opinion. This means that even when there is enough committed agents to cause consensus on opinion $A$, from our first critical fraction of agents, the stickiness dominates the system and allows for only a few remaining members to hold unique opinions. However when $c_A>2s-1$, the active steady-state returns as seen in Fig. 9. 

\begin{figure}[h]
\centering
\includegraphics[width=0.5\textwidth]{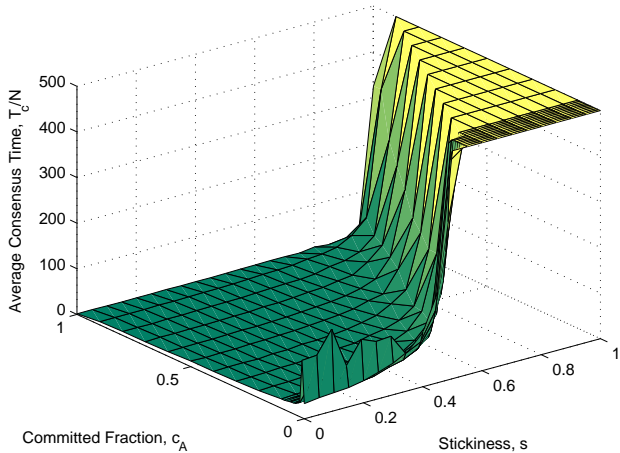}
\caption{Average ratio of time to consensus, $T_c$, to the network size, $N$, as a function of stickiness, $s$, and the fraction, $c_A$, of nodes committed to opinion $A$. Lighter color implies longer consensus times.}
\end{figure}

Figure 10 shows the consensus time normalized by the networks size, $T_c/N$, taken over twenty runs. Like in Figure 4, values of $T_c/N = 500$ represent a lower bound of the normalized consensus time, as this value was used to stop the simulations if the consensus was not reached by then. As shown in Figure 7, for values $\frac{1}{2} < s \leq 1$, a triangular $AB$ dominant region exists where neither consensus to opinions $A$ nor to $B$ occur. For lower values of $s$, the transition between $A$ and $B$ dominance to an active steady-state occurs for lower fractions of committed agents then predicted. Despite this minor difference, when the fraction of agents committed to opinion $A$ is greater than the threshold value, an active steady-state with majority of $A$ opinion holders occurs. Conversely, for small values of the fraction of committed agents, opinions $A$ and $B$ coexist without either achieving lasting majority. 

\section{Summary}
The proposed models are developed using a mean field on a complete graph. Since the mean actions are considered, the model accurately describes the network when a large population exists. When smaller groups are considered, these findings do not to take into account the random fluctuations and the stochastic stabilities causing discrepancies in the models' accuracy. In a social network, it is not necessarily true that individuals are familiar with and talk to every other member of network. In such a case, an incomplete graph should be used instead of a complete graph. The linearization used to check stability of these models is accurate when non-degenerate stabilities are considered. However, when degenerate stabilities are found, it is not sufficient to consider only linear terms and higher order terms need to be analyzed. 

By introducing the parameters for propensity and stickiness into the naming game, we have uncovered both a drift in equilibrium points as well as the occurrence of a new stability point (Table V). The propensity of speakers in the system is intended to capture some external influence such as media or opinions supported by a government leader which introduces some bias (or barrier) in willingness of members of the network to share an opposing opinion with others. Such barrier or bias requires a larger than usual majority of opinion holders against the external influence to create a drift towards the opposing opinion consensus. If this larger threshold is not met, the network will reach a consensus on opinion supported by external influence. The stickiness model will lead to an inversion in the stability of equilibrium points. This inversion allows for a consensus like state in the mixed opinion; the system has a majority of members holding this opinion. If holders of mixed opinion are reluctant to accept any unique opinion, stubbornly sticking to their current one, it makes it less likely that any unique opinion will dominate the network. Such stubborn individuals act like social mediators, preventing dominance by any unique opinion. As the stickiness to mixed opinion by its holders becomes stronger and stronger, the change of opinions and consequently a global consensus become less and less likely. The two-parameter model integrates effects of the introduced two parameters; such integration increases system complexity, some aspects of which warrant further study. 

\begin{table}
\footnotesize
\caption{Effects of variations to the naming game}
\begin{tabular}{|p{1.5cm}|p{2.2cm}|p{4.5cm}|}
\hline
Parameter & Range of Parameters & Effects on Linear Stability\\ \hline
Propensity, p & 0 $\leq p < \frac{1}{2}$ & Stronger drift towards consensus on A than on B.\\
& $\frac{1}{2} < p \leq 0$ & Stronger drift towards consensus on B than on A.\\ \hline
Stickiness, s & $0 \leq s < \frac{1}{2}$ & Same stability as naming game with drift towards both A and B. \\
& $\frac{1}{2} < s \leq 1$ & Inversions of stability of naming game, with drift towards 'consensus' at mixed opinion.\\ \hline
\end{tabular}
\end{table}

The addition of committed agents to the stickiness model has been shown to hold similar properties to that of the naming game solely with committed agents. The stickiness model with committed agents has two bifurcations which create barrier to reaching an active steady-state at which the committed opinion becomes a global consensus. This model has properties similar to the stickiness model. When stickiness is weak, a barrier appears near the opinion without a committed majority, inhibiting any drift towards consensus on the opinion held by the committed agents. Instead, a consensus-like state appears in the opposing opinion, as seen in the standard model for the naming game. As the fraction of committed agents reaches critical point, their influence overcomes the opinions of the populace and drives the system to consensus. 

When stickiness is strong, a barrier appears closer to the committed opinion critical fraction but it is buffered by the holders of mixed opinion, as seen in the model with only stickiness present. Hence, the stickiness increases the critical fraction of committed agents necessary to overcome the mixed opinion holders' barrier. With the fraction of committed agents below the critical threshold, the state of the system becomes a consensus-like state with the majority of the network members holding mixed opinion. As the fraction of committed agents increases to critical point, the pull of the committed minority overcomes the power of stickiness, creating a strong drift towards consensus on the committed opinion. Overall, the stickiness of holders strongly controls the dynamics of the committed agent model.

\begin{acknowledgments}
This work was supported in part by the Army Research Laboratory under Cooperative Agreement Number W911NF-09-2-0053, by the Army Research Office Grants No. W911NF-09-1-0254 and W911NF-12-1-0546, by the Office of Naval Research Grant No. N00014-09-1-0607 and by the EU's 7FP Grant Agreement No. 316097. The views and conclusions contained in this document are those of the authors and should not be interpreted as representing the official policies, either expressed or implied, of the Army Research Laboratory or the U.S. Government.
\end{acknowledgments}

\end{document}